\documentclass[preprint]{elsarticle}
\usepackage{amssymb,amsmath}
\usepackage{graphicx}
\usepackage{array}
\usepackage{ifthen}

\usepackage{figwidth}

\newlength{\mywidth}

\newcommand{\reffig}[1]{Fig.~\ref{#1}}
\newcommand{\reffigs}[1]{Figs.~\ref{#1}}
\newcommand{\reftable}[1]{Table~\ref{#1}}

\newcommand{\ansatz}[3]{\begin{array}{c}#1\\#2\\#3\end{array}}

\begin{document}

\ifthenelse{\equal{\Qclass}{revtex4}}{
\title{The topological susceptibility from grand canonical simulations in the interacting instanton liquid model: strongly associating fluids and biased Monte Carlo.}

\author{Olivier Wantz}
\email[Electronic address: ]{O.Wantz@damtp.cam.ac.uk}
\affiliation{Department of Applied Mathematics and Theoretical Physics,
Centre for Mathematical Sciences,\\ University of Cambridge,
Wilberforce Road, Cambridge CB3 0WA, United Kingdom}
}{}
\ifthenelse{\equal{\Qclass}{elsarticle}}{
\title{The topological susceptibility from grand canonical simulations in the interacting instanton liquid model: strongly associating fluids and biased Monte Carlo.}
\author[damtp]{Olivier Wantz}
\ead{O.Wantz@damtp.cam.ac.uk}
\address[damtp]{Department of Applied Mathematics and Theoretical Physics,
Centre for Mathematical Sciences,\\ University of Cambridge,
Wilberforce Road, Cambridge CB3 0WA, United Kingdom}
}{}

\bibliographystyle{plain}

\begin{abstract}
This is the second in a series of papers that investigates the topological susceptibility in the interacting instanton liquid model (IILM); it deals with the technical issues relating to the Monte Carlo simulations that are specific to finite temperature. The IILM reduces field theory to a molecular dynamics description, and for `physical' quark masses the system behaves like a strongly associating fluid. We will argue that this is a generic feature for very light Dirac quarks in a non-trivial background, described in the semi-classical approach. To get rid of unnecessary complications, we will present the ideas of biased Monte Carlo, and implement the transition probabilities, for a toy model.
\end{abstract}

\maketitle

\section{Introduction}
\label{sec:introduction}

In a recent paper \cite{wantz:iilm:1}, we have performed simulations in the interacting instanton liquid model (IILM) at zero temperature. In this and the next paper \cite{wantz:iilm:3} we will address finite temperature simulations. Using the light `physical' quark masses obtained in \cite{wantz:iilm:1}, we have found that the Monte Carlo techniques that worked well at zero temperature become rather inefficient at finite temperature. The reason lies in the formation of instanton--anti-instanton pairs. They provide the mechanism for chiral symmetry restoration within the IILM  \cite{ilgenfritz:shuryak:chiral:symmetry:restoration:iilm}, \cite{ilgenfritz:shuryak:quark:correlations:chiral:transition}, \cite{schaefer:shuryak:verbaarschot:chiral:phase:transition:molecules}: because the quark wavefunctions become localised on those pairs, the Dirac eigenvalue spectrum $\rho(\lambda)$ has no support at zero virtuality; according to the Banks-Casher relation \cite{banks:casher:chiral_symmetry},
\begin{equation}
 \langle \bar{q}q\rangle = -\pi \rho(\lambda=0)\,,
\end{equation}
chiral symmetry is restored. This is in contrast to zero temperature where the quark wavefunctions are delocalised and the eigenvalue spectrum extents down to zero virtuality. The localisation of the Dirac zero modes at finite temperature is potentially troublesome for ordinary Monte Carlo techniques because it induces short-ranged interactions between instantons.

In another numerical study \cite{schaefer:shuryak:iilm} of the IILM at finite temperature, no technical problems were encountered because the quark mass parameters were large enough for ordinary Monte Carlo to work well. But as the masses decrease, the interactions become stronger and random sampling starts to run into trouble. Furthermore, in the deconfined phase screening sets in: the temperature fluctuations obstruct the formation of coherent field configurations that exceed the screening length, so that instanton sizes are cutoff at $\rho \lesssim 1/T$. Since the interactions in the IILM follow from the overlap of the instanton profiles, smaller sizes lead to shorter-ranged forces between pairs.

The strong attraction is a generic feature of the IILM at finite temperature. Normalising the low frequency determinant to the dilute gas, the quark effective interaction is of the form
\begin{equation}
 S^q = -\ln \det \left( \mathbb{I} + T^2/M^2 \right)\,. \label{eq:quark:interaction}
\end{equation}
Depending on the topological charge, the determinant is over the subspace spanned by the $N_I$ or $N_A$ zero modes: $T^2=T^\dagger T$ if $N_I>N_A$, and $T^2=TT^\dagger$ otherwise. The matrix $M^2$ is diagonal and of the same dimension as $T^2$; its non-vanishing elements are given by the squared quark mass. Crucially, it is an attractive interaction because the determinant is bounded from below by unity. It is clear from the above expression that the quark interaction becomes ever stronger as quark masses decrease.

The feature of strong and short-ranged interactions is thus not confined to the specific case of the trivial holonomy calorons \cite{harrington:shepard:caloron} that we have been studying, but is a general characteristic of non-trivial backgrounds; in particular, it will play a role for the newly found non-trivial holonomy calorons \cite{kraan:baal:caloron:I,kraan:baal:caloron:II} and \cite{lee:lu:caloron}. The latter might be the correct degrees of freedom to explain the confinement/deconfinement phase transition \cite{gerhold:ilgenfritz:mueller_preussker:kvbll:gas:confinement}, \cite{diakoniv:gromov:petrov:slizovskiy:caloron:su_2:weight}, the lack of which is a major shortcoming of the current IILM.

Systems with strong and localised interactions have been investigated for a long time in chemical engineering and computational chemistry, and are known to present computational challenges. They run under the name of strongly associating, or ionic, fluids. The technical problems that Monte Carlo methods face are two-fold:
\begin{itemize}
 \item The small relative volume of attraction makes ordinary Monte Carlo updates miss them most of the time.
 \item Once a pair is formed, ordinary Monte Carlo moves can get stuck in these configurations because of the large energy difference.
\end{itemize}
From an algorithmic point of view, it means that the acceptance rates to reach or leave the regions of phase space corresponding to instanton--anti-instanton molecules are very low. This would not be a problem if we had an infinite amount of computer time, but, in practice, it leads to very long autocorrelation times. This, in turn, leads to strong dependence on initial conditions and, in the extreme case of very poor mixing between correlated and uncorrelated states, the system might stay fixed in one of those `phases' during the available computer time. For all practical purposes, we lose ergodicity because we end up with samples that are either devoid or dominated by pairs, none of which is a representative sample.

The solution is to develop algorithms which explicitly sample the attraction centres and are able to break up pairs with high acceptance rates. To sample these volumes, we need to, first, construct these regions and, then, to define a measure over them. This is a non-trivial geometrical problem; it also depends very strongly on the specific problem. Considering the dipole character of the instanton interactions, this seems like a daunting task.

In recent years, general purpose algorithms have been developed which do not rely on an accurate construction of the union of all attraction volumes but on the different ways that a given, simple interaction box can be reached, e.g.\ the Unbonding--Bonding algorithm \cite{wierzchowski:kofke:associating_fluids}. In its original form, it has been given for a canonical ensemble. We will adapt it to our needs for grand canonical simulations.

As argued above, the quark interactions will generically lead to strong interactions for small quark masses, and any non-trivial background will lead to screening effects at finite temperature. To avoid unnecessary complications and background dependent features, e.g.\ the colour-orientation dependence for calorons, we will use a toy model that mimics the interactions of (\ref{eq:quark:interaction}).

In section \ref{sec:markov:chain:monte:carlo} we will quickly review the idea behind Markov chain Monte Carlo. We will then discuss biased Monte Carlo in section \ref{sec:biased:monte:carlo}, and set up the framework for dealing with strong and short-ranged interactions in the grand canonical ensemble. Finally, we will present a toy model in section \ref{sec:monte:carlo:toymodel}, and benchmark the biased moves from section \ref{sec:biased:monte:carlo} within that setting.

\section{Markov Chain Monte Carlo}
\label{sec:markov:chain:monte:carlo}

The phase space for a general statistical mechanical system will be very high-dimensional for a large number of particles. Usual integration rules over a grid are not well adapted to evaluate the defining partition function because, for such high dimensional integrals, the number of grid points would be overwhelmingly large.

Markov chain Monte Carlo techniques are the method of choice\footnote{If the integrand is positive definite} because they can `find' the relevant regions that dominate the partition function. It is achieved by preferentially sweeping those parts of parameter space that have a large measure. It can be shown \cite{norris:markov:chains} that $p^n_{ij} \to p^{\mathrm{eq}}_j$, where $p_{ij}$ is the transition probability and $p^{\mathrm{eq}}$ is the unique invariant equilibrium distribution
\begin{equation}
 \sum_i p^{\mathrm{eq}}_i p_{ij}=p^{\mathrm{eq}}_j\,. \label{eq:invariant:distribution}
\end{equation}
 So-called ergodic theorems relate ensemble averages to time averages of paths through phase space, i.e.\
\begin{equation}
 \frac{1}{N}\sum_{i=1}^N f(X_i) \to \langle f \rangle = \sum_{i \in I} p^{\mathrm{eq}}_i f(x_i)\,,
\end{equation}
with a specific path given by the Markov chain $\{X_n\}$.

In our case, we know $p^{\mathrm{eq}}$, essentially the integrand of the partition function, and we want to construct $p_{ij}$ that converge to this equilibrium distribution. Using the convergence theorems, this is a well posed problem if we demand that $p^{\mathrm{eq}}$ is invariant for $p_{ij}$. It is not hard to see that (\ref{eq:invariant:distribution}) is fulfilled if we impose the stronger condition of detailed balance,
\begin{equation}
 p_i^\mathrm{eq} p_{ij} = p_j^\mathrm{eq} p_{ji}\,. \label{eq:detailed:balance}
\end{equation}
This deceptively simple looking equation lies at the heart of all Monte Carlo simulations. Note that the $p_{ij}$'s are not unique and that there is a large amount of freedom in choosing them. This redundancy can be used to accelerate convergence.

If there are different paths that connect states $i$ and $j$, $p_{ij}=\sum_a^N p^a_{ij}$, (\ref{eq:detailed:balance}) can be met by imposing the stronger condition
\begin{equation}
 p_i^\mathrm{eq} p^a_{ij} = p_j^\mathrm{eq} p^a_{ji}\,. \label{eq:super:detailed:balance}
\end{equation}
This is sometimes called super-detailed-balance \cite{frenkel:smit:understanting_molecular_simulation}. If there are very many different paths, $N\gg 1$, the implementation can become rather tedious, and super-detailed-balance might be the only viable option. Even if there are a manageable number of paths, the book-keeping needed to relate them might be overwhelming.

It is important to note that the transition probability is really the product of the proposal probability $\mathcal{P}_{ij}$ and the acceptance probability $\mathcal{A}_{ij}$. This observation lies at the heart of biased Monte Carlo techniques: we can tweak the proposal probabilities to increase acceptance rates. Typically, this leads to asymmetric transition probabilities. The latter can, however, also be found in ordinary Monte Carlo: consider for instance a simple insertion/deletion step for grand canonical simulations; the proposal probability for an insertion corresponds to the probability to place the particle within the simulation box, whereas for a deletion it gives the probability to choose a particle already in the box. Clearly, these will be different in general.

We will focus on the Metropolis algorithm \cite{frenkel:smit:understanting_molecular_simulation}, which defines the acceptance probability by
\begin{equation}
 \mathcal{A}_{ij} = \min \left( 1, \frac{p^\mathrm{eq}_j}{p^\mathrm{eq}_i} \frac{\mathcal{P}_{ji}}{\mathcal{P}_{ij}} \right)\,. \label{eq:metropolis}
\end{equation}
It is not hard to see that it satisfies (\ref{eq:detailed:balance}).

\section{Biased Monte Carlo}
\label{sec:biased:monte:carlo}

But for the simplest statistical systems, ordinary Monte Carlo techniques can become rather inefficient, if not downright inadequate: inefficiency manifests itself by long autocorrelation times, and inadequacy stems from an inability to sample the relevant regions accurately. For finite computer time, and hence finite-sized samples, we basically lose ergodicity.

By `ordinary' Monte Carlo moves, we mean random sampling. A typical example is updating the position of a particle: it is straightforward to implement such a move by adding a random increment to the particle's current position. This random increment is typically drawn from a fixed volume with uniform measure. Thus, $\mathcal{P}_{ij}$ is constant for all updates and cancels in (\ref{eq:metropolis}).

For strong and short-ranged interaction, such a random increment will be very inefficient. Either we choose it so small that we can sample the interaction region, in which case very many sweeps are needed to move through phase space; this leads to very long autocorrelation times. Or we choose a large enough increment, so as to sweep quickly through phase space, but thereby missing the interaction regions most of the time; the sample will most likely not be representative. Furthermore, the acceptance probability for random sampling is typically given by the ratio of two Boltzmann factors, and does only depend on the energy difference. Due to the strong interaction, the acceptance probability will be very low for moves that attempt to leave the attraction centres. The Markov chain can become trapped in these energy-dominated configurations, which leads again to a non-representative sample.

The way out is to use importance sampling. It is designed to sample those parts of phase space that dominate the partition function. It might help to get a rough criterion for importance sampling. Remember that in our case Monte Carlo methods try to evaluate the integral given by the partition function. If the integrand is peaked so strongly that the exact integral is well approximated by the integral over some localised patches, the algorithm will need to preferentially sample these parts of phase space. Thus, random sampling will be inefficient if
\begin{equation}
 V < \Delta V \exp(-H(\Delta V))\,, \label{eq:pair:criterium}
\end{equation}
where $V$ is the volume of the simulation box, and $\Delta V \ll V$ is the small region where the interaction is very strong.

As mentioned in the introduction, in chemical engineering and computational chemistry the issues of low acceptance rates have long been known: they are very low in polymer physics simulations, due to conformational obstructions; or in simulations of ionic fluids, due to strong short-ranged interactions. The latter is of immediate interest to us because the IILM at finite temperature displays the same characteristics. To treat such systems correctly, new algorithms based on biased Monte Carlo have been developed \cite{visco:kofke:associating_fluids}. We can take advantage of these well tested techniques in IILM simulations at finite temperature.

Recently, efficient and general purpose algorithms have been developed \cite{chen:siepmann:associating_fluids} \cite{wierzchowski:kofke:associating_fluids}. In these algorithms, the focus is not on an accurate construction of the union of all the interaction regions, which is a difficult, and problem-dependent, geometric task, but on the individual interaction regions and all the possible routes that lead to the same final state. In \cite{chen:siepmann:associating_fluids}, the algorithm is further simplified by using super-detailed-balance (\ref{eq:super:detailed:balance}), whereas the algorithm in \cite{wierzchowski:kofke:associating_fluids} does not rely on this stronger condition, and was shown to converge faster. We use the latter scheme, the Unbonding--Bonding algorithm (UB).

The UB algorithm starts by defining a bonding region. It does not necessarily need to be the exact physical bonding region, but a strong departure from it will not be very useful. A list is made of those instantons that are in at least one bonding region of an anti-instanton and the same is done for the anti-instantons. There are $N^B_I$ ($N^B_A$) bonded instantons (anti-instantons). $N^B_I(i_I)$ is the number of anti-instantons that instanton $i_I$ is bonded to, and analogously for $N^B_B(i_A)$. In what follows, $i_I$ means instanton $i_I$, but can also stand for the state the instanton $i_I$ is in; unprimed quantities are evaluated before the move, whereas primed ones denote the same quantity after the move.

We will now focus on the UB for instantons; the case for anti-instantons is then obvious. The bonding move consists of choosing uniformly an instanton and an anti-instanton, and placing the instanton in the bonding region of the anti-instanton with flat measure. The unbonding move consists of choosing one of the bonded instantons and to place it randomly in the simulation box; again both steps are performed with a flat measure for the bonded instantons and the simulation box respectively. This leads to the following transition probabilities
\begin{align}
 \mathcal{P}^B_{i_I(i'_I,i_A)} & = \frac{1}{N_I}\frac{1}{N_A} \frac{1}{V_{i_A}}\,,\\
 \mathcal{P}^U_{ii'} & = \frac{1}{N^B_I} \frac{1}{V}\,,
\end{align}
where $(i_I,i_A)$ is a bonded instanton--anti-instantion pair, and $V_{i_A}$ is the bonding region of anti-instanton $i_A$. The UB algorithm now adds up all possible routes that lead to the same final state $i'$, bonding and unbonding; super-detailed-balance is, however, used with respect to the unbiased displacement move. The forward and backward proposal probabilities are then given by
\begin{align}
 \mathcal{P}_{i_I i'_I} & = \sum_{i_A}^{N'^B_I(i_I)} \mathcal{P}^B_{i_I(i'_I,i_A)} + \delta^B_{i_I} \mathcal{P}^U_{i_I i'_I}\,,\\
 \mathcal{P}_{i'_I i_I} & = \sum_{i_A}^{N^B_I(i_I)} \mathcal{P}^B_{i'_I(i_I,i_A)} + \delta^B_{i'_I} \mathcal{P}^U_{i'_I i_I}\,,
\end{align}
with $\delta^B_i=1$ if $i$ is bonded and $\delta^B_i=0$ otherwise. We assume that bonding and unbonding moves have the same a-priori-probability, $1/2$, which we omitted because it cancels out anyway.

Since we want to perform grand canonical simulations, we also need insertion and deletion moves. The unbiased moves are given by
\begin{align}
 \mathcal{P}^\mathrm{ub}_{N_I,N_I+1} & = \frac{1}{V}\,,\\
 \mathcal{P}^\mathrm{ub}_{N_I+1,N_I} & = \frac{1}{N_I+1}\,.
\end{align}
The corresponding biased insertions and deletions will be constructed along the lines of the UB algorithm, either by placing the instanton $i_I$ into the bonding region of an anti-instanton, or by removing the bonded instanton $i_I$. More precisely, insertions consist of choosing an anti-instanton $i_A$, placing the instanton $i_I$ uniformly within the bonding box $V_{i_A}$ and finally summing over all possible anti-instantons that could have been chosen to reach this same final state; to delete an instanton, we choose uniformly from the list of bonded instantons $N_I^B$. In formulas,
\begin{align}
 \mathcal{P}^\mathrm{b}_{N_I,N_I+1} & = \sum_{i_A}^{N'^B_I(i_I)} \frac{1}{N_A} \frac{1}{V_{i_A}}\,,\\
 \mathcal{P}^\mathrm{b}_{N_I+1,N_I} & = \frac{\delta^B_{i}}{N'^B_I}\,.
\end{align}
In this case, it does not produce much overhead to combine the biased and unbiased moves. More importantly, we found that acceptance rates could be boosted upon combining biased and unbiased insertions/deletions. To add them up, we need to specify the relative weights, the a-priori-probability for biased updates $p_b$. The full insertion/deletion proposal probabilities are
\begin{align}
 \mathcal{P}_{N_I,N_I+1} & = p_b \mathcal{P}^\mathrm{b}_{N_I,N_I+1} + (1-p_b)\mathcal{P}^\mathrm{ub}_{N_I,N_I+1}\,,\\
 \mathcal{P}_{N_I+1,N_I} & = p_b \mathcal{P}^\mathrm{b}_{N_I+1,N_I} + (1-p_b)\mathcal{P}^\mathrm{ub}_{N_I+1,N_I}\,.
\end{align}
To reiterate, a similar factor for the a-priori-probabilities of canonical moves was tacitly omitted before because we chose to follow the original implementation of the UB algorithm, which does not mix biased and unbiased particle updates: its use of super-detailed-balance implies that the a-priori-probabilities give an overall multiplicative factor that drops out.

Other than good mixing between bonded and unbonded structures, cluster moves have been argued to be important to accelerate convergence towards the equilibrium distribution \cite{orkoulas:panagiotopoulos:associating_fluids}. For the IILM, we assume that instanton--anti-instanton pairs are dominating in this respect. We therefore augment our list of single particle moves by pair-displacements and pair-insertions and -deletions.

Whereas we made sure to have a uniform distribution among the $N^B_I$ bonded instantons for the UB-type moves discussed so far, regardless of whether they were bonded many times, we naturally demand uniformity in the number of pairs $N_P$ for the pair-moves.

The pair-displacements consist of collective translations of the pair, by an increment $\delta \in v_C$, and of internal displacements, $\delta \in v_I$, of one of the pair's constituents. The latter are rejected if the displaced instanton leaves the bonding box. The proposal probabilities for these moves are given by
\begin{align}
 \mathcal{P}^C & = \frac{1}{N_P} \frac{1}{v_C}\,,\\
 \mathcal{P}^I & = \frac{1}{N_P} \frac{1}{v_I}\,,
\end{align}
We use super-detailed-balance for the internal displacement because the acceptance rates could not be boosted by including unbiased and/or UB moves; the additional overhead, required to go beyond super-detailed-balance, is thus not justified.

The pair-insertions and -deletions, set up without recourse to super-detailed-balance, are built from biased and unbiased moves, whose probabilities are summed up in the end. A biased pair-insertion consists of placing, uniformly, either an instanton $i_I$ or an anti-instanton $i_A$ in the simulation box; the partner is then positioned randomly within the bonding box. Note that the probabilities for both possibilities are added up. The unbiased pair-insertion consists of placing, uniformly, both an instanton $i_I$ and an anti-instanton $i_A$ in the simulation box. For a biased pair-deletion we select an instanton and anti-instanton uniformly from the pairs $N_P$, whereas for an unbiased pair-deletion we select an instanton and anti-instanton randomly from $N_I$ and $N_A$ respectively. The final proposal probabilities are then given by
\begin{align}
 \mathcal{P}_{N,N+1} & = p'_b \delta^P_{i_I i_A} \left( \frac{1}{2} \frac{1}{V} \frac{1}{V_{i_I}} + \frac{1}{2} \frac{1}{V} \frac{1}{V_{i_A}} \right) + (1-p'_b) \frac{1}{V} \frac{1}{V}\,,\\
 \mathcal{P}_{N+1,N} & = p'_b \frac{\delta^P_{i_I i_A}}{N'_P} + (1-p'_b) \frac{1}{N'_I}\frac{1}{N'_A}\,.
\end{align}
The factor $\delta^P_{i_I i_A}$ ensures that the biased contribution to the proposal probability is added only if $i_I$ and $i_A$ are paired, i.e.\ $\delta^P_{i_I i_A}=1$ if instanton $i_I$ and anti-instanton $i_A$ are paired and $\delta^P_{i_I i_A}=0$ otherwise. The a-priori-probabilities $p'_b$ can be chosen to tune the acceptance rates even further.

Given these different proposal probabilities, together with the trivial case of unbiased displacements, we can compute the acceptance probability through (\ref{eq:metropolis}).

Finally, we need to specify the increment volumes $v_i$ and the bonding boxes $V_i$, and we have to decide upon the different a-priori-probabilities $p_b$, $p'_b$ and $p_i$. The latter give the probability to perform a specific update and have, so far, been omitted because they will always cancel in the acceptance probabilities. These choices will depend on the problem at hand, and some fine-tuning runs cannot be avoided to set these parameters.

In practice, we have finite-sized samples, and the outcome of the simulations will depend on these parameters. The differences will vanish with increasing sample size; this observation provides a straightforward means to test whether the biased updates have been implemented correctly. For computer-intensive simulations, large samples are prohibitive, and it is important that the dependence is rather weak if the algorithms are to be useful.

\section{Toy Model}
\label{sec:monte:carlo:toymodel}

Now that we have set up the update moves using biased Monte Carlo techniques, we will test their performance compared to random sampling. We will use a two-dimensional toy model that nevertheless will mimic the IILM. The advantage of the toy model is that it will be computationally very cheap. Also, we can adjust the parameters freely to get a rough estimate for those regions in parameter space where pair formation will be important and, hence, biased Monte Carlo essential.

\begin{figure}[tbp]
\begin{center}
 \includegraphics[width=\figwidth,clip=true,trim=0mm 0mm 15mm 10mm]{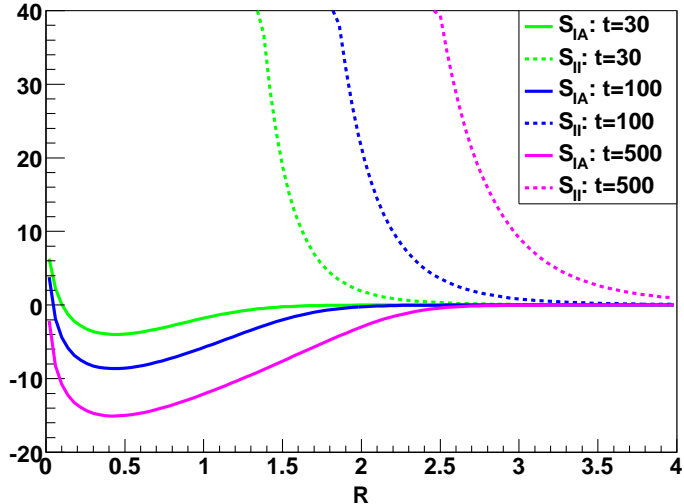}
\end{center}
 \caption{The solid lines represent the interaction for an instanton--anti-instanton pair. The repulsion for like-charged pairs is represented by dashed lines. It is chosen so strong that clusters larger than simple pairs are strongly disfavoured. We see that as the temperature $t$ is raised the attractive well deepens. In this graph $\alpha=2$.}\label{fig:toymodel:interactions}
\end{figure}

The interactions in the IILM have the following form
\begin{align}
 S_{IA}^\mathrm{int} &= \ln \left(1+\frac{1}{R^2} \right) - \frac{1}{1+R^2} - N_f \ln\left( 1+e^{-2\pi R} R^2 \frac{T^2}{m^2} \right)\,, \label{eq:toymodel:ia:interaction} \\
 S_{II}^\mathrm{int} &= \ln \left(1+\frac{1}{R^2} \right) - \frac{1}{1+R^2}\,.
\end{align}
The logarithmic repulsion in the pair separation $R$, given here in units of the inverse temperature $T$, is typical for the IILM\footnote{In the ratio ansatz, see \cite{wantz:iilm:1}.}. It combines with the rational function to produce the large separation decay. The instanton--anti-instanton pairs feel an additional attraction due to the quark wavefunction overlaps. Their $R^2$ dependence follows from `strong' overlaps, whereas the exponential describes the large separation behaviour. For our purposes, the product of both will be a good enough approximation. The quark mass is given by $m$, the number of flavours by $N_f$. As for the IILM, $S_{II}$ and $S_{IA}$ have the same fall-off behaviour.

In two dimensions, this isotropic interaction results in a rather dense ensemble because the repulsion is mild and the attractive wells are comparatively deep. More importantly, these interactions will favour large clusters of instantons and anti-instantons. Such a behaviour does not correspond to the situation that we expect from the IILM, namely the formation of rather isolated pairs. In the IILM, the interaction is orientation dependent, and large conglomerates of instantons cannot form because most of the relative orientations within the cluster would be repulsive. To achieve this same effect of pair formation with our toy model, we add an extra `entropic' repulsion term to $S^{II}_\mathrm{int}$.

The exponential decay for the quark overlaps starts for $ e^{-2\pi R} t^2 \approx 1$; we have set $t=T/m$, i.e.\ the temperature in units of the quark mass. For separations smaller than $R_0 = \ln t/\pi$, the attraction wells are rather deep, and we choose the repulsion to become strong in order to obstruct the formation of large clusters. The functional form of this `entropic' repulsion is rather unimportant, and for simplicity we choose it to be given by
\begin{equation}
 S^\mathrm{ent}_{II} = \left(\frac{\alpha R_0}{R}\right)^8\,.
\end{equation}
It turns out that choosing $\alpha=2-3$ is sufficient to prevent cluster formation. For dilute ensembles, the precise value is rather irrelevant.

Now that we have fixed the interactions, see also \reffig{fig:toymodel:interactions}, we define our system by the following partition function
\begin{align}
 Z &= \sum_{N_I, N_A}^\infty \int dx^{N_I+N_A} \frac{d^{N_I}}{N_I!} \frac{d^{N_A}}{N_A!} \exp\left(-S_\mathrm{int}(x)\right)\,, \label{eq:partition:function:toy:model}\\
 S_\mathrm{int} &= \sum_{i<j}^{N_I} S_{ij}^\mathrm{int} + S_{ij}^\mathrm{ent} + \sum_{i<j}^{N_A} S_{ij}^\mathrm{int} + S_{ij}^\mathrm{ent} + \sum_{i_I i_A}^{N_I,N_A} S_{i_I i_A}^\mathrm{int}\,,
\end{align}
where the free parameters of the model are $d$, $t$ and the simulation box $V$. In the full model, instantons have a size $\rho$, and quantum effects are such that small sizes are disfavoured, $d(\rho)=\rho^\beta$ with $\beta>0$ \cite{thooft:instanton:fluctuations}; the free parameter $d$ plays that role.

Neglecting interactions, $d$ determines the density of the ensemble,
\begin{equation}
 \frac{\langle N \rangle}{V}=2d\,. \label{eq:toymodel:dilute:gas}
\end{equation}
This is the dilute gas result. Once interactions are included, the system will be more dilute if the total interaction is repulsive and less dilute otherwise; this is adjusted by the parameter $t$ which captures the combined effects of temperature and quark masses, see (\ref{eq:toymodel:ia:interaction}).

\begin{figure}[tbp]
\begin{center}
 \includegraphics[width=0.85\figwidth]{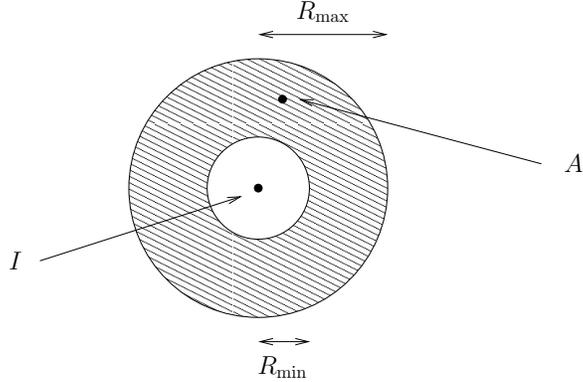}
\end{center}
 \caption{For the biased Monte Carlo moves to work well, the bonding box should be adapted to the interactions. Isotropy suggests the best form to be in the shape of an annulus. The shaded region corresponds to the bonding box. A priori $R_\mathrm{min}$ and $R_\mathrm{max}$ are free parameters that can be fine-tuned to achieve good mixing, i.e.\ low autocorrelation times.}\label{fig:toymodel:bondingbox}
\end{figure}

In order to use the biased Monte Carlo framework set up in section \ref{sec:biased:monte:carlo}, we need to decide on the bonding box to use. Given that the interaction is isotropic, we will choose an annulus as bonding box, see \reffig{fig:toymodel:bondingbox}. It is natural to fix the free parameters $R_\mathrm{min}$ and $R_\mathrm{max}$ so as to achieve low autocorrelation times.

\begin{figure}[tbp]
\begin{center}
 \includegraphics[width=\figwidth,clip=true,trim=0mm 0mm 15mm 10mm]{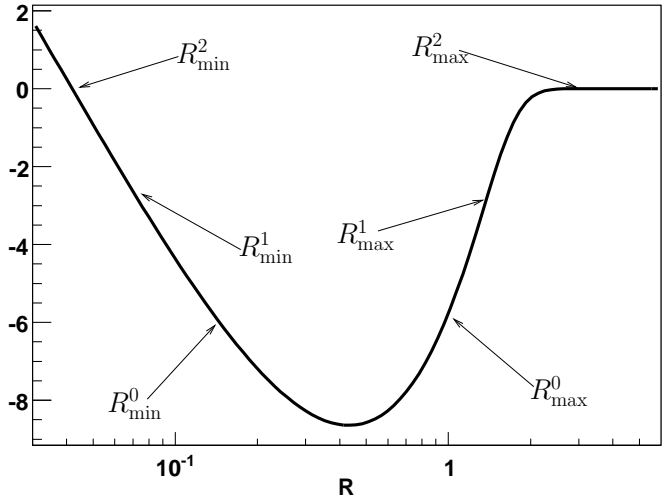}
\end{center}
 \caption{We choose three different sets for $(R^i_\mathrm{min},R^i_\mathrm{max})$. A small interval that samples the strong interaction region, a `natural' one that sample most of the attractive well and a larger box that extends beyond what would naturally look like the bonding region.}\label{fig:toymodel:bondingboxes}
\end{figure}

We perform biased simulations for three sets of bonding boxes, and compare them with unbiased runs. The natural box size extends to $R_\mathrm{max}=R_0$, the point at which the quark interaction starts to vanish exponentially. The corresponding lower edge $R_\mathrm{min}$ is determined by $S_{IA}^\mathrm{int}(R_\mathrm{min}) \approx S_{IA}^\mathrm{int}(R_\mathrm{max})$. Given the relatively large values for $t$, we can solve this approximately by
\begin{equation}
 R_\mathrm{min} \approx \exp\left(-\frac{1}{6} S_{IA}^\mathrm{int}(R_\mathrm{max}) - \frac{2}{3} \ln t \right)\,.
\end{equation}
We also choose a larger interval with $R_\mathrm{max}=2 R_0$ and a smaller one that samples predominantly the very strong interaction region. In the latter case, we choose $R_\mathrm{min}$ and $R_\mathrm{max}$ according to $S_{IA}^\mathrm{int}(R_\mathrm{max}) \approx S_{IA}^\mathrm{int}(R_\mathrm{min}) \approx \frac{1}{3}\min_R S_{IA}^\mathrm{int}$. A typical setup is displayed in \reffig{fig:toymodel:bondingboxes}.

\begin{figure}[tbp]
\begin{center}
 \includegraphics[width=\figwidth,clip=true,trim=0mm 0mm 15mm 10mm]{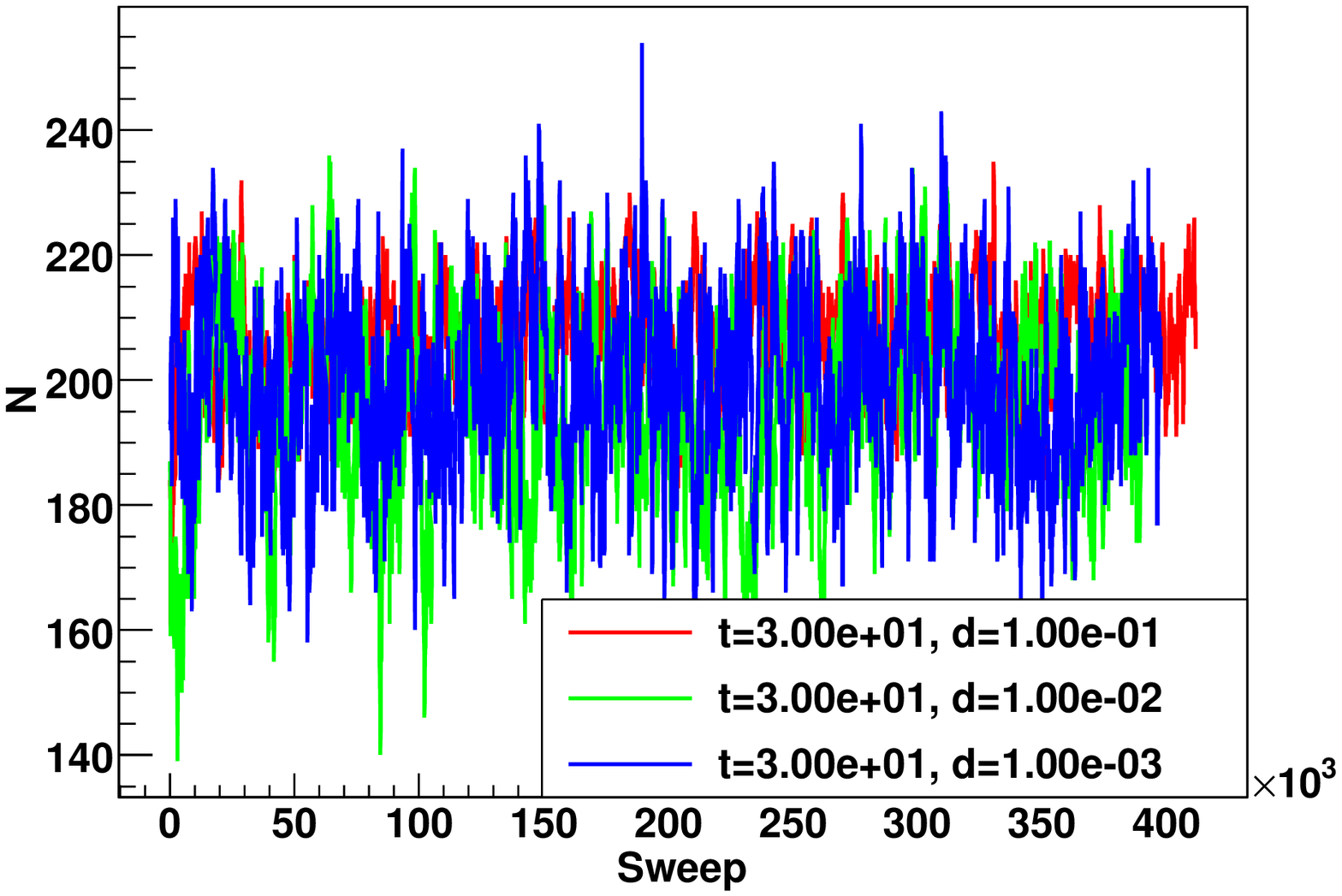}
 \includegraphics[width=\figwidth,clip=true,trim=0mm 0mm 15mm 10mm]{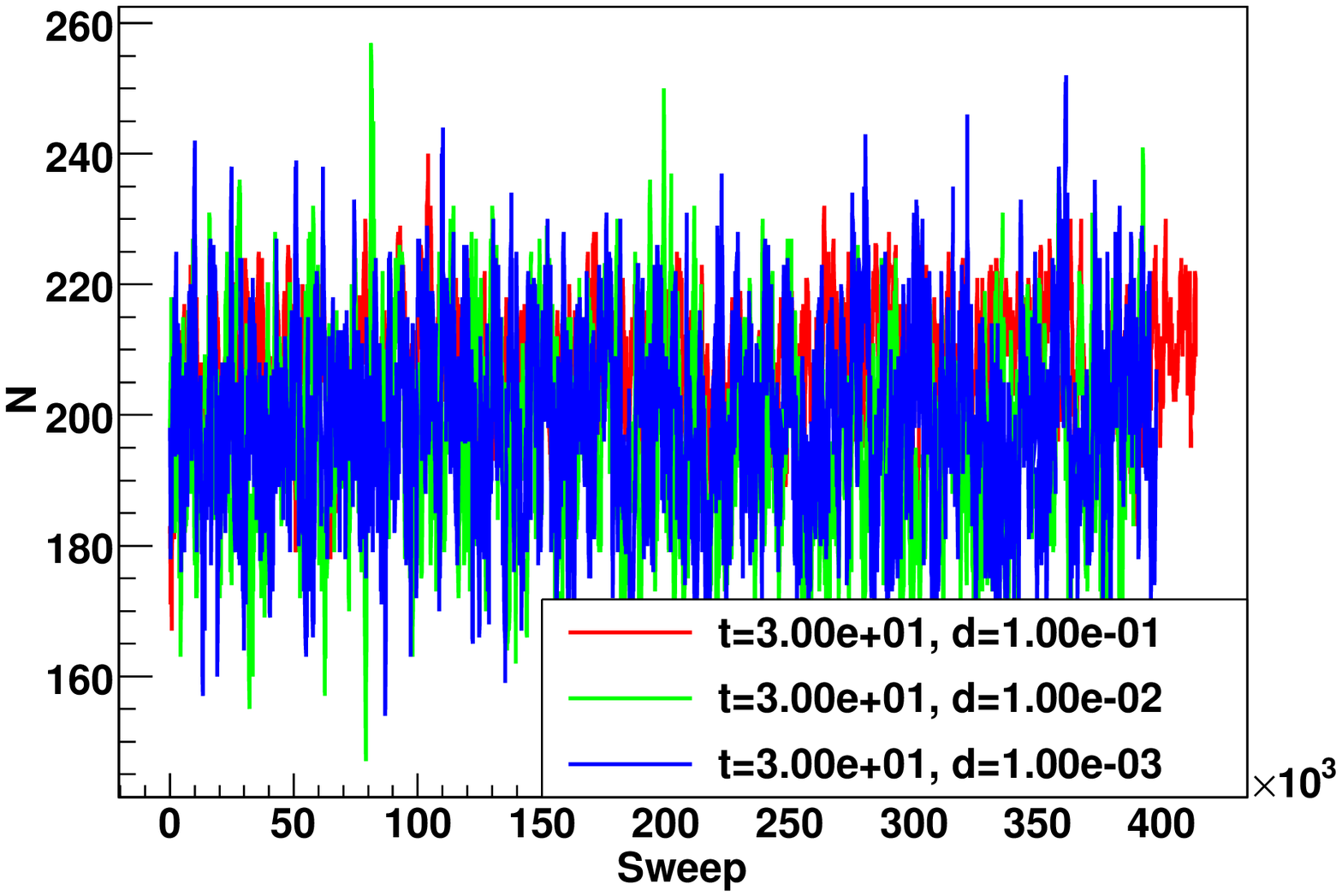}
\end{center}
 \caption{Choosing a rather small parameter for $t$, i.e.\ a shallow attraction well, the biased (bottom) and unbiased (top) simulations have thermalised over the same time scale. The overhead that importance sampling introduces is not necessary provided that the autocorrelation times are similar. In the present case we found that this is indeed so.}\label{fig:toymodel:thermalization:t0}
\end{figure}

\begin{figure}[tbp]
\begin{center}
 \includegraphics[width=\figwidth,clip=true,trim=0mm 0mm 15mm 10mm]{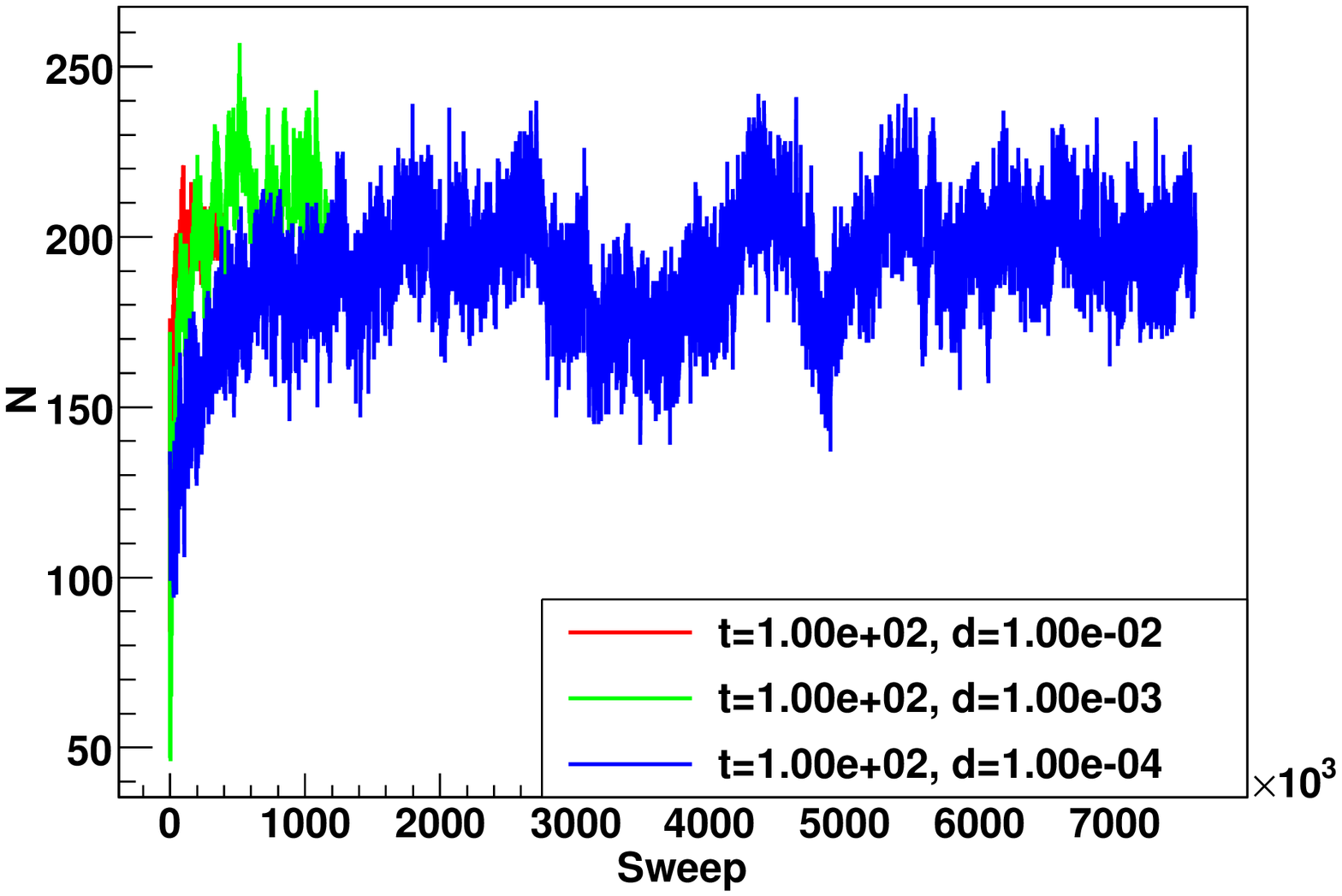}
 \includegraphics[width=\figwidth,clip=true,trim=0mm 0mm 15mm 10mm]{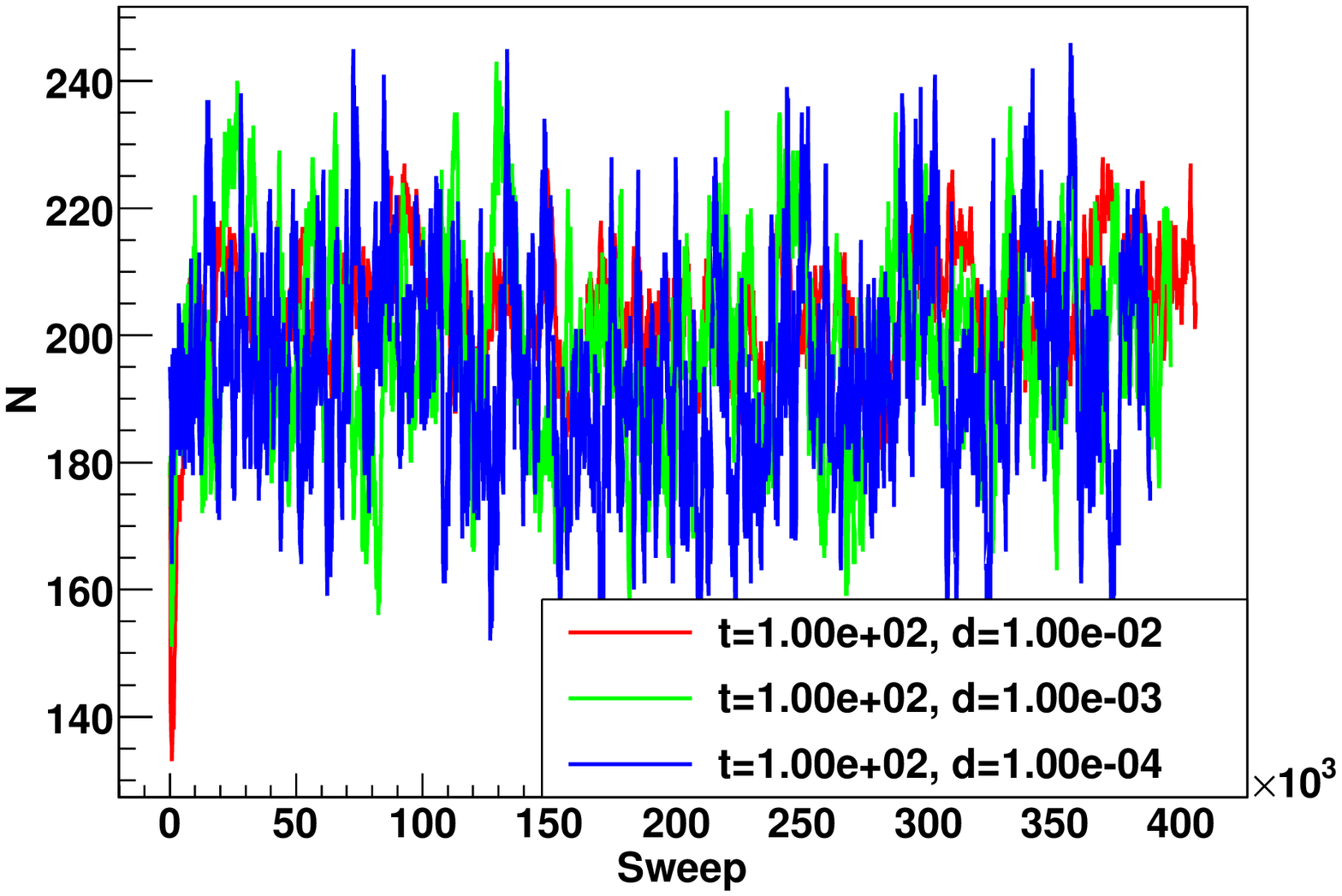}
\end{center}
 \caption{As the interaction well is deepened, i.e.\ for larger $t$, unbiased Monte Carlo algorithms (top) start to run into trouble. The thermalisation process takes longer and longer as we increase the dilution of the system, i.e.\ by decreasing $d$. Importance sampling (bottom) thermalises much faster: notice the rather large difference in the number of sweeps it takes to reach equilibrium for the blue ($d=10^{-4}$) curve.}\label{fig:toymodel:thermalization:t1}
\end{figure}

Before we start the simulations, it is worth mentioning one more technical point, related to the choice for the equilibrium probability density. We claim that it is given by
\begin{equation}
 p^\mathrm{eq} = \frac{d^{N_I} d^{N_A} e^{-S_\mathrm{int}}}{Z}\,. \label{eq:equilibrium}
\end{equation}
We want to put attention to the fact that in (\ref{eq:equilibrium}) we neglect the Boltzmann-counting factors for indistinguishable particles. We would like to argue that they are an artifact. These terms are introduced so as to allow for an unconstrained integration region, $x^a_i \in [0,L^a]$, instead of the complicated multi-dimensional region given by $\{x^a_i | x^a_1 < x^a_2 \cdots < x^a_{N_j}\}$, with $a$ labelling a component. The result of the integrations is corrected by dividing by the factorials because $\cup_\sigma \{x^a_i | x^a_{\sigma(1)} < x^a_{\sigma(2)} \cdots < x^a_{\sigma(N_j)}\} = [0,L^a]^{N_j}$, where $\sigma$ is a permutation. Therefore, the Boltzmann-counting factors are not really part of the weight \cite{kofke:private_communication:indistinguishability}. A different, but equivalent, point of view is to keep the factorials in the weight and to sum the equilibrium probability, with the factorials included, over all permutations $N_i !$ that correspond to mere relabellings of the particles in the initial state \cite{rull:jackson:smit:gibbs_ensemble}. The permutations cancel off the factorials, and we are left again with (\ref{eq:equilibrium}). The transition probabilities $P_{ij}$ are similarly summed over all initial and final state permutations, and the resulting factor of $N_i ! N_j!$ trivially cancels out of (\ref{eq:detailed:balance}). The simplest way to understand the need to sum over permutations, lack of which will violate detailed balance, is to realise that, taking indistinguishability at face value, the states are really equivalence classes\footnote{The group by which we factor out is the permutation group}. Since we actually use a specific representative but the result should be independent of it, we need to sum over all members of the given equivalence class. The upshot is that, for all practical purposes, we treat the different particles as distinguishable!

The parameter $t$ is set to the values given in \reffig{fig:toymodel:interactions}. The dilution parameter $d$ is chosen small enough so that we end up with a rather dilute system because the biased Monte Carlo moves, set up in section \ref{sec:biased:monte:carlo}, will not work well for dense ensembles. We are interested in how the system behaves if $d$ is further decreased. We can envisage three different equilibrium states: either the ensemble will be dominated by pairs, or equilibrates in a mixture of paired and unpaired instantons, with roughly equal weight, or settles in an uncorrelated state with no pairs.

\begin{table}[tbp]
\begin{center}
\begin{tabular}{c|c|c|c}
MC & $d$ & $\langle N \rangle$ & $\xi$ \\ \hline\hline
&&&\\[-0.35cm]
biased & $\ansatz{10^{-2}}{10^{-3}}{10^{-4}}$ & $\ansatz{202.4(2)}{199.7(4)}{193.1(4)}$ & $\ansatz{7940}{7302}{2857}$\\ \hline
&&&\\[-0.35cm]
unbiased & $\ansatz{10^{-2}}{10^{-3}}{10^{-4}}$ & $\ansatz{202.0(1)}{200.2(1)}{196.8(2)}$ & $\ansatz{41058}{75115}{140683}$\\ \hline
\end{tabular}
\end{center}
\caption{We estimate the autocorrelation times $\xi$ for the instanton number $N$ as a function of the Monte Carlo algorithm and the dilution parameter $d$. The temperature/mass parameter is $t=100$, i.e.\ it corresponds to the data in \reffig{fig:toymodel:thermalization:t1}. We clearly see that as the density of the system is lowered, i.e.\ by decreasing $d$, the autocorrelation times for unbiased simulations become very large compared to the unbiased ones.}\label{table:toymodel:autocorrelation:t1}
\end{table}

\begin{figure}[tbp]
\begin{center}
 \includegraphics[width=\figwidth,clip=true,trim=0mm 0mm 15mm 10mm]{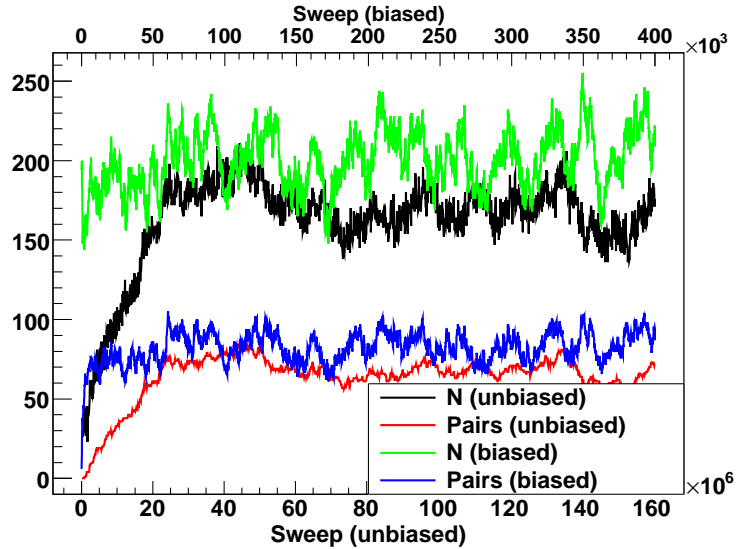}
\end{center}
 \caption{The long thermalisation time is directly correlated with the slow process of forming pairs in the unbiased case. The correlation between pairs and instanton number is also seen in the biased simulations; here thermalisation is fast (note the different scale, i.e.\ the upper axis) because the algorithm explicitly samples the possible bonding sites.}\label{fig:toymodel:thermalization:NvP:hot}
\end{figure}

For the simulations that follow, we use the `natural' bonding box, $R_\mathrm{max}=R_0$. We always use hot initial conditions, i.e.\ place the instantons and anti-instantons randomly throughout the volume. As we can see from \reffig{fig:toymodel:thermalization:t0}, the smallest value of $t$ leads to equivalent equilibrium states for both biased and unbiased Monte Carlo. However, as we increase $t$, and lower $d$ in order to maintain a dilute ensemble, ordinary Monte Carlo needs a very long time to reach equilibrium compared to biased Monte Carlo, see \reffig{fig:toymodel:thermalization:t1}. Since we start with a random distribution, the long thermalisation process can be attributed to the fact that it takes longer and longer to locate the attraction centres as the system becomes ever more dilute. The correlation between the number of instantons $N$ and the number of pairs $P$ is clearly visible in \reffig{fig:toymodel:thermalization:NvP:hot}.

\begin{figure}[tbp]
\begin{center}
 \includegraphics[width=0.6\figwidth]{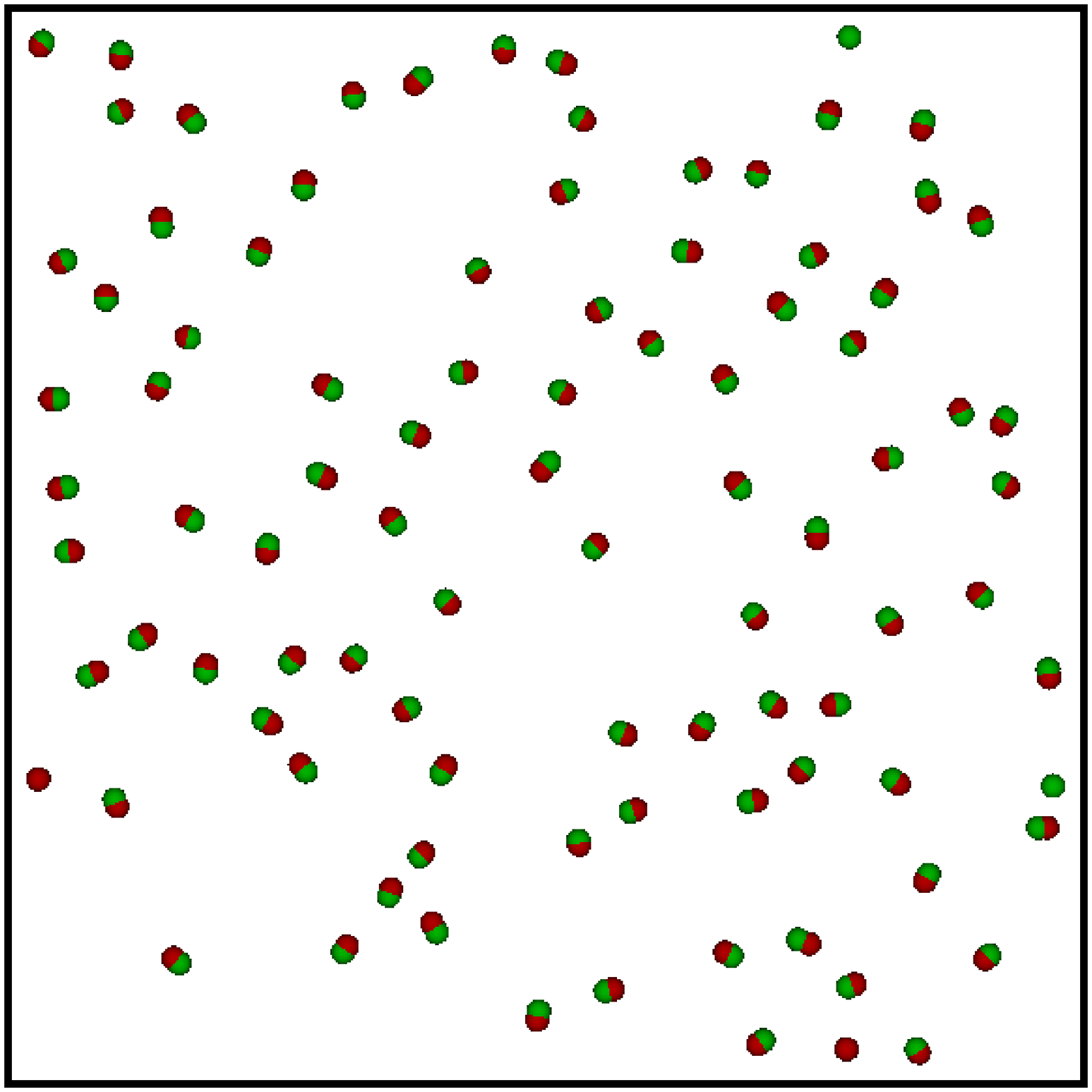}
 \includegraphics[width=0.6\figwidth]{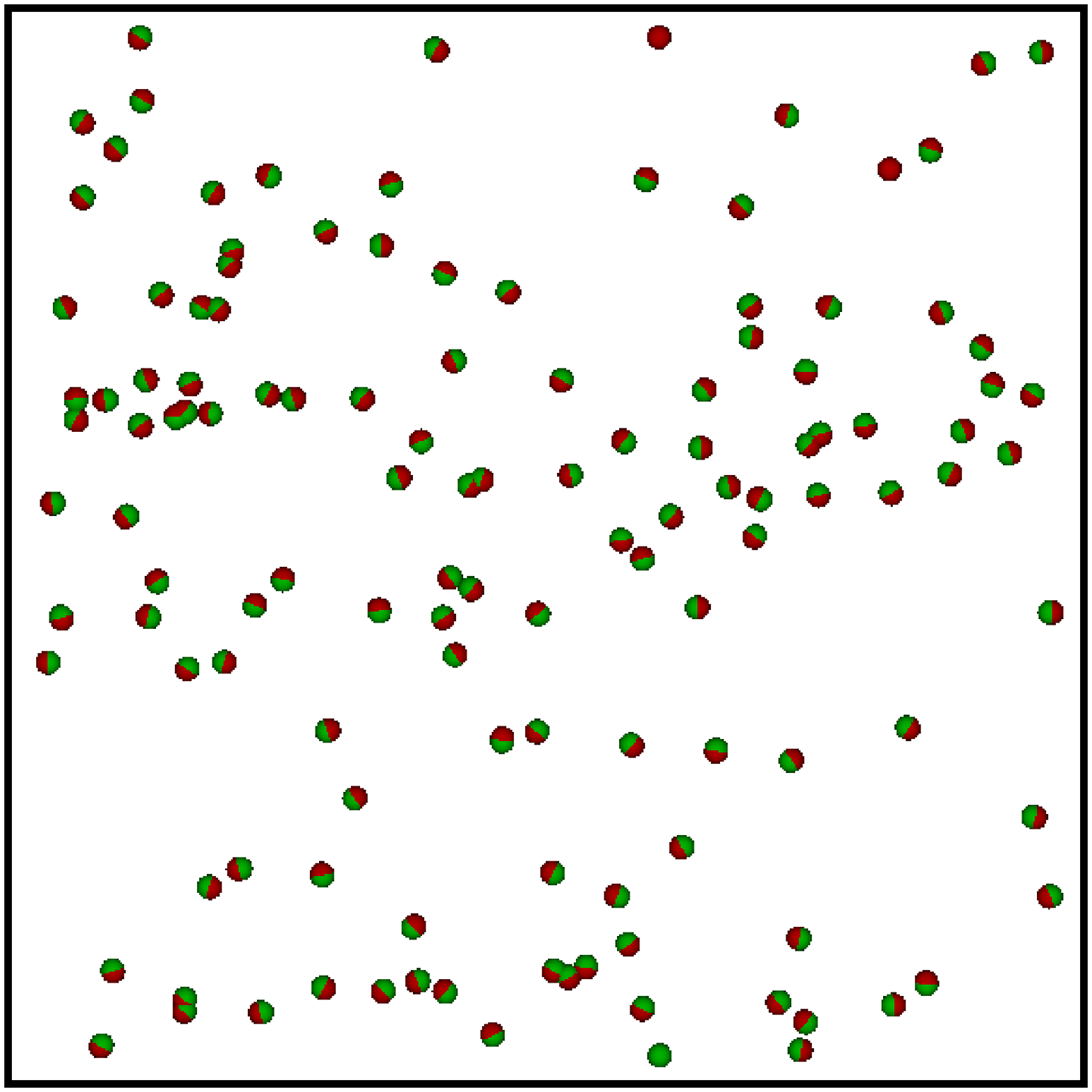}
 \includegraphics[width=0.6\figwidth]{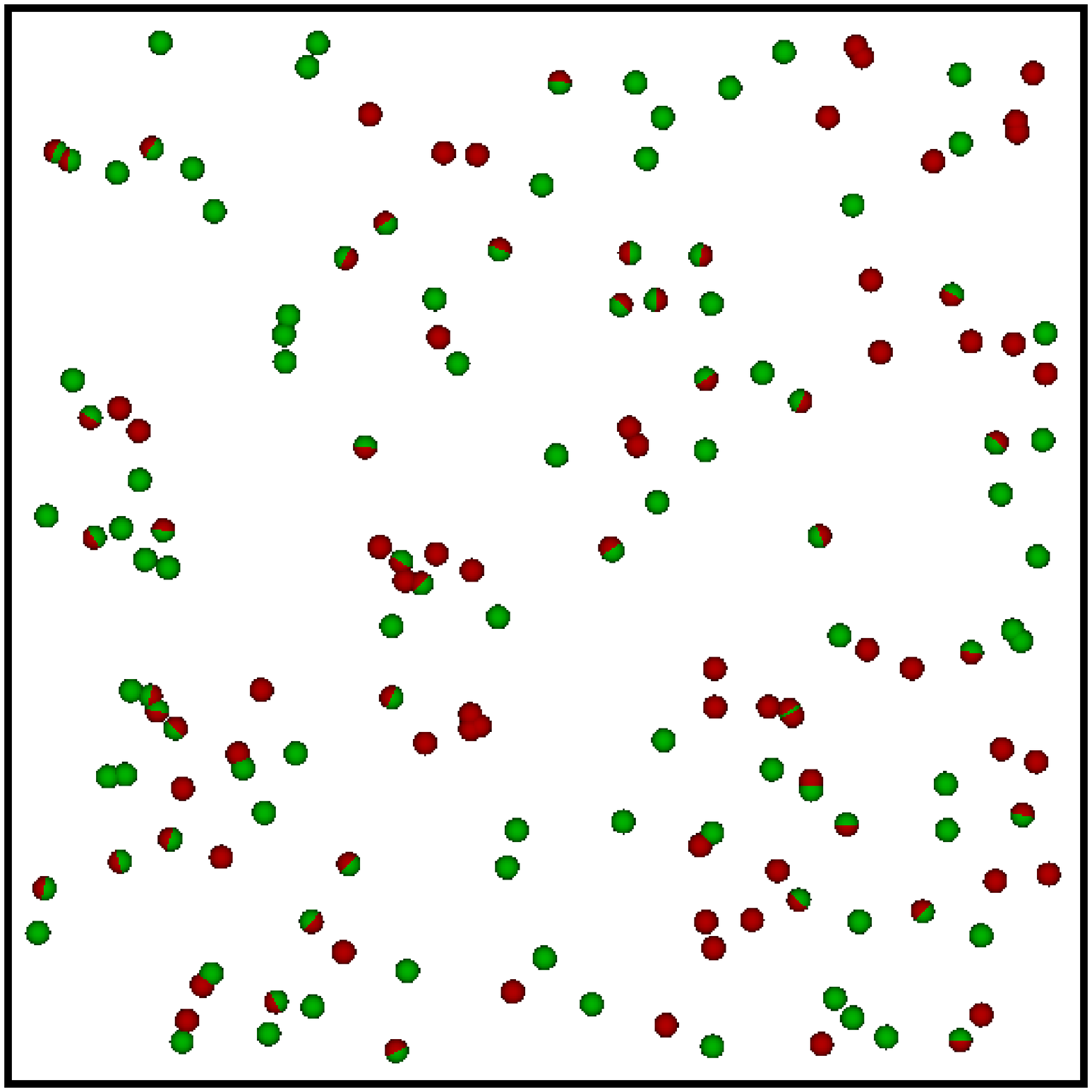}
 \includegraphics[width=0.6\figwidth]{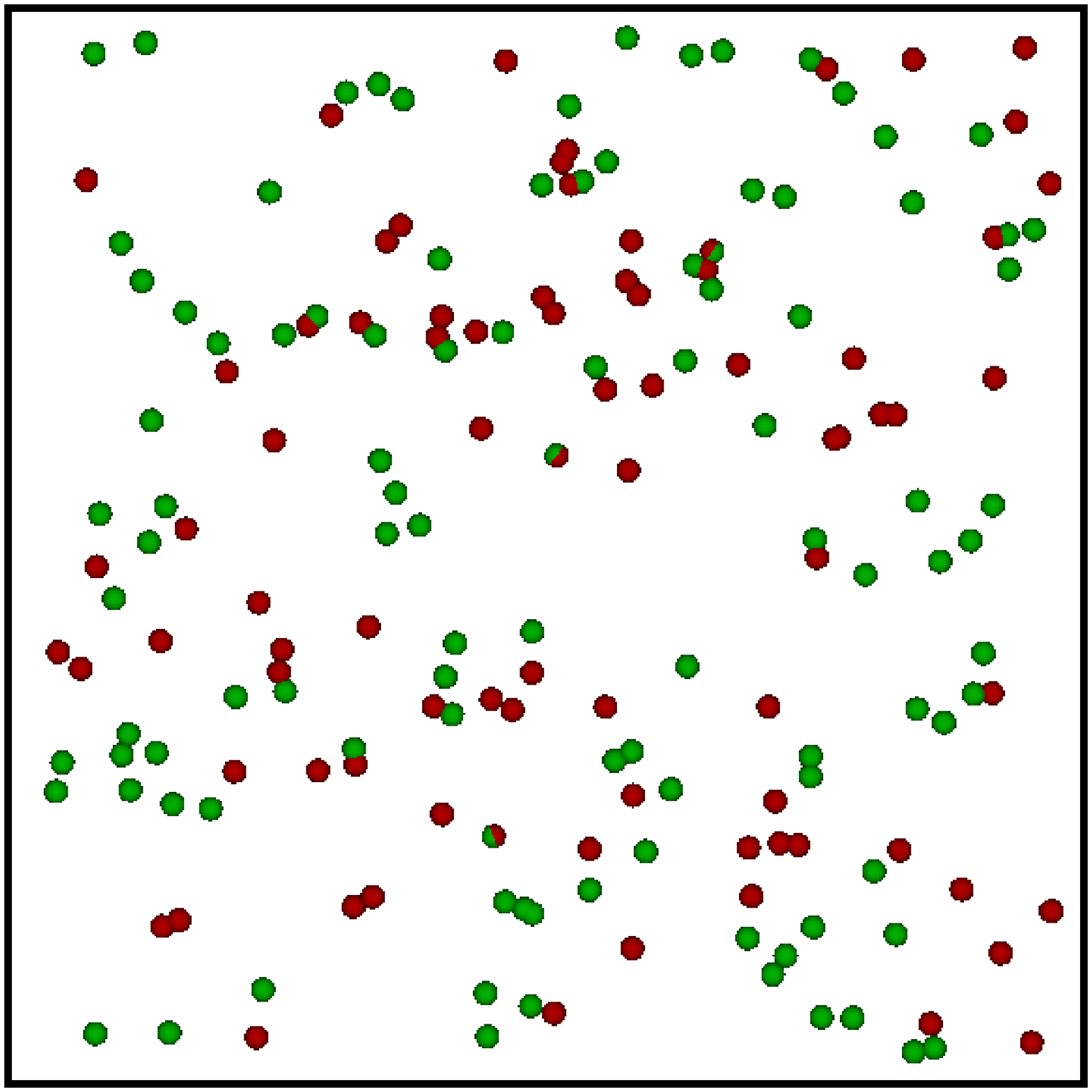}
\end{center}
 \caption{The degree of dilution increases from left to right, top to bottom; the temperature-mass parameter is fixed at $t=300$. As $d$ is decreased fewer instantons are paired up with anti-instantons. For the last box the system has equilibrated in a random state.}\label{fig:toymodel:evolution:with:d}
\end{figure}

After equilibration, we can estimate autocorrelation times. We find that the autocorrelation time in the instanton number $N$ is much larger for random sampling simulations, see \reftable{table:toymodel:autocorrelation:t1}. The reason is that random insertions and deletions are suppressed: insertions  because they generally do not form pairs, and deletions because they most likely try to break up pairs; this leads to low acceptance probabilities. A useful quantity is the autocorrelation time per instanton. It does not change with the system size, i.e.\ it is intensive; equivalently, autocorrelation times are extensive, i.e.\ scale linearly with the system size, which in our case can be parametrised by the number of instantons. Note that the autocorrelation times for the biased simulations actually drop in contrast to the unbiased ones. The reason is that as $d$ decreases the system consists of a gas of instanton--anti-instanton pairs with less and less interactions, so that the updates on pairs will become more and more efficient whereas acceptance rates for ordinary Monte Carlo drop ever more\footnote{Note that in the limit $d \to 0$, with $t=\mathrm{const}$, the system behaves as a dilute gas of instantons and that ordinary insertions and deletions will work fine again. See \reffig{fig:toymodel:evolution:with:d} for more details.}. As the quark mass parameter $t$ is increased, keeping $d$ fixed, the autocorrelation times increase as well; the reason is that the system becomes denser, so that the probability is higher for the proposal configurations to be rejected due to repulsive interactions\footnote{Note that in the limit $t \to \infty$, with $d=\mathrm{const}$, the system becomes too dense to be treated with the methods developed in this paper.}.

The toy-model is computationally rather cheap, which allows for very long thermalisation sweeps. During such a long equilibration process, helped by the low dimensional configuration space, random sampling does eventually form sufficient pairs to converge to the same state as importance sampling. In higher dimensional simulations, as for instance in the IILM, the interaction regions have much lower `entropy'\footnote{By which we mean that $\Delta V/V \ll 1$, with $\Delta V$ the bonding volume.}, and it will be harder for ordinary Monte Carlo simulations to find the regions of phase space that lead to pair formation. Also, realistic systems are computationally more expensive\footnote{For instance, the IILM in the unquenched case needs to evaluate determinants, a time-consuming task.}, and long runs are prohibitive. Biased Monte Carlo techniques become unavoidable in these situations.

\begin{figure}[tbp]
\begin{center}
 \includegraphics[width=\figwidth,clip=true,trim=0mm 0mm 15mm 10mm]{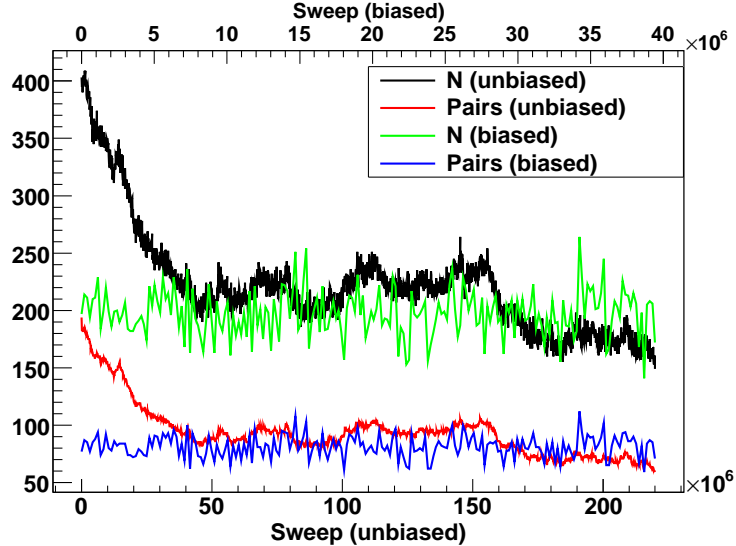}
\end{center}
 \caption{Random sampling is not well suited to deal with cold initial conditions: we started the thermalisation run with a configuration made up of $200$ pairs. Random sampling has trouble to break up the excess pairs. The inefficiency can be traced back to the fact that the algorithm depends solely on the energy difference,which is big when destroying pairs in a `naive' way. This leads to low acceptance probabilities and hence long thermalisation sweeps. The biased simulations equilibrate very quickly because the algorithm is tailored to deal with bonded instantons.}\label{fig:toymodel:thermalization:NvP:cold}
\end{figure}

The hot initial condition manifests itself by a rapid drop in instanton numbers early in the thermalisation process, due to the lack of pairs; this can even be seen in the case of biased simulations, although the algorithm creates pair much more quickly, see \reffig{fig:toymodel:thermalization:t1}. We expect that cold initial conditions, i.e.\ starting off with pairs, will also be problematic for random sampling techniques because they rely solely on the energy difference between states: in see \reffig{fig:toymodel:thermalization:NvP:cold} we examine how ordinary Monte Carlo copes with an ensemble with excess pairs.

From \reffigs{fig:toymodel:thermalization:t1} and \ref{fig:toymodel:thermalization:NvP:cold} it is clear that biased Monte Carlo techniques vastly outperform random sampling. The computational advantage follows from our knowledge of the structures that can form, e.g.\ instanton--anti-instanton molecules in the IILM \cite{ilgenfritz:shuryak:chiral:symmetry:restoration:iilm}, and the ability to construct algorithms that take full advantage of that knowledge.

So far we have restricted the simulations to a single bonding box, which is a free parameter a priori. We have run every data set with the two other bonding boxes, see \reffig{fig:toymodel:bondingboxes}. We have found that the results agree on the $2\sigma$ level. In \reftable{table:toymodel:bondingboxes}, we show such data for one specific point in parameters space $\{d,t\}$. This particular set, with large $t$ and very small $d$, is probably representative for more realistic simulations in that the unbiased simulations have failed to converge to the true equilibrium state; we have lost ergodicity because those regions of phase space that correspond to bonded pairs is not sampled adequately.

The weak dependence on the bonding box allows us to dismiss systematic effects, and we can base our choice entirely on achieving low autocorrelation times. The data presented in \reftable{table:toymodel:bondingboxes} would suggest that the two smallest bonding boxes are equivalent but the bulk of the computations suggest the lowest autocorrelation times are achieved with the smallest bonding box.

\begin{table}[tbp]
\begin{center}
\begin{tabular}{c|c|c|c}
Bonding Box $(R_0) $& $\langle N \rangle $ & $\langle Q^2 \rangle $ & $ \langle S_{\mathrm{int}} \rangle $ \\ \hline\hline
$(0.102328,0.661723)$&$194.6(4)$&$134(4)$&$-2.323(5)$\\
&[1998]&[1790]&[1083] \\ \hline
$(0.0457066,1)$&$194.7(4)$&$133(3)$&$-2.328(7)$\\
&[1907]&[1212]&[1812] \\ \hline
$(0.0313807,2)$&$195.5(3)$&$137(3)$&$-2.340(7)$\\
&[4629]&[2607]&[5475] \\ \hline
unbiased&$131.8(3)$&$129(5)$&$-0.4(6) \,10^{-5}$\\
&[887]&[322]&[545] \\ \hline
\end{tabular}
\end{center}
\caption{This data is from a simulation with $t=500$ and $d=10^{-7}$, and for these parameters the unbiased Monte Carlo simulation equilibrates at a substantially lower value for the total number of instantons $N$. Note also the low interaction, from which we conclude that no pairs have formed. The results for the different bonding boxes agree well at the $2\sigma$ level. The autocorrelations are given in square brackets. The two smallest boxes give rather similar autocorrelation times but we have found that in most of the runs the smallest box leads to the fastest convergence.}\label{table:toymodel:bondingboxes}
\end{table}

It is interesting to see how the ensemble behaves as we tune the two free parameters $d$ and $t$ separately. We have found that the system will always evolve to a random state with a negligible amount of pairs upon decreasing $d$ for any fixed $t$. This is as it should be because small $d$ favour a dilute system: from (\ref{eq:toymodel:dilute:gas}) we see that $V$ increases for decreasing $d$ and constant number of instantons; the strong interaction region $\Delta V$ is unaffected by $d$, and the inequality (\ref{eq:pair:criterium}) will be violated for large enough $V$; thus, random sampling is efficient and the ensemble equilibrates in an uncorrelated state. We illustrate this in \reffig{fig:toymodel:evolution:with:d}.

\begin{figure}[tbp]
\begin{center}
 \includegraphics[width=\figwidth,clip=true,trim=0mm 0mm 15mm 10mm]{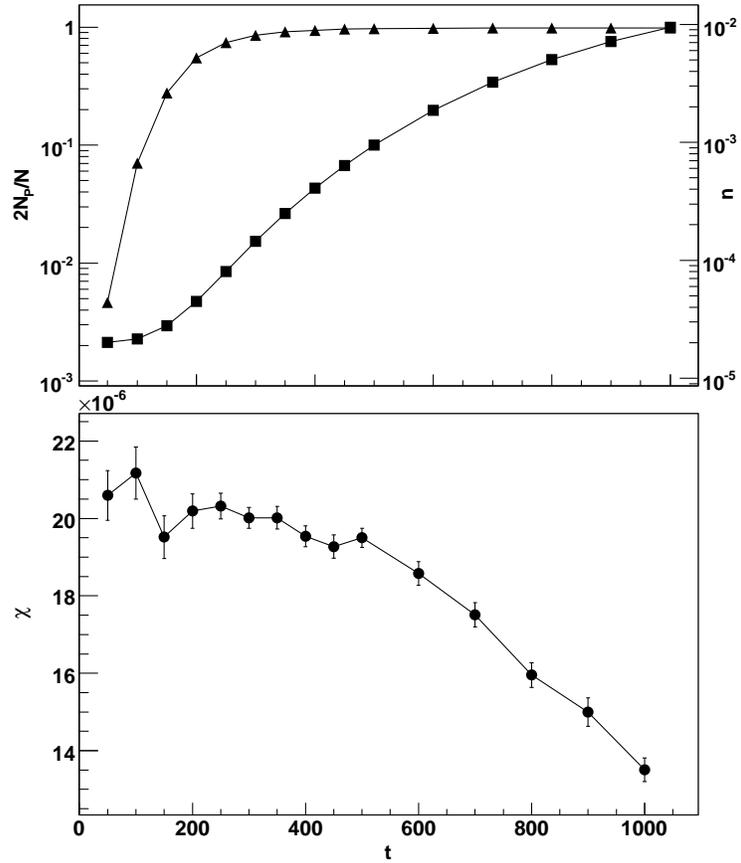}
\end{center}
 \caption{As the interaction strength $t$ is increased, the proportion of pairs to instantons increases until basically all instantons are in pairs. Pairs do not contribute to topological fluctuations, reducing the topological susceptibility. However, the instanton density increases rather strongly with $t$, resulting in a much denser ensemble. This implies a larger topological susceptibility, because the fluctuations per unit volume increase. Both effects are clearly antagonistic. In this case, it turns out that the topological susceptibility decreases, although rather slowly.}\label{fig:toymodel:nP_n_Q2}
\end{figure}

Similarly, we find that for fixed dilution $d$ and increasingly large interaction strength $t$ the system tends to favour energy-dominated configurations. We therefore expect that the topological susceptibility should drop to zero since pairs do not contribute to the charge fluctuations. The instanton density, on the other hand, increases because the energy-dominated configurations favour a denser ensemble. This in turn implies that the topological susceptibility will be boosted because the topological fluctuations per unit volume have increased. Both effects work in opposite directions, but it turns out that the topological susceptibility decreases overall, although rather slowly. This is demonstrated in \reffig{fig:toymodel:nP_n_Q2}. In particular, the topological susceptibility is lower than the dilute gas approximation would suggest.

In QCD, the level of dilution and the interaction strength change simultaneously, such as to increase both with increasing temperature. Therefore, we cannot draw any direct conclusions from this work, other than to note that with increasing fermion number the screening, and hence the dilution, decreases whereas the fermionic interactions increase. It is therefore not clear a priori whether in QCD the IILM will remain in a highly correlated molecular phase after the chiral phase transition.

\section{Conclusions}

We have argued that, quite generically, light quarks in non-trivial backgrounds will induce strong and short-ranged interactions at finite temperature. The Dirac operator is positive definite and bounded, and leads to an attraction between instantons that becomes stronger with decreasing quark mass. Plasma effects at finite temperature constrain the spatial extent of the non-trivial backgrounds to be below the screening length; since the quark interactions are induced by the overlap of quark wavefunctions, centred on top of the classical gauge fields, it follows that the interactions become more and more short-ranged as the temperature rises.

Systems with strong and short-ranged interactions are known as strongly associating fluids in the literature of computational chemistry and chemical engineering. These fluids are hard to simulate and need special techniques. Using the concepts of the general purpose Unbonding--Bonding algorithm, we extended the biased scheme to grand canonical simulations. Following suggestions from the literature, we augmented the Monte Carlo steps by pair moves. The input from previous studies of the IILM is crucial as these have identified the instanton--anti-instanton pairs to be important at finite temperature. This knowledge has allowed us to implement biased Monte Carlo techniques that specifically address the problems faced with random sampling.

The defining characteristic of the system is its strong and short-ranged interactions. We therefore have decided to test importance sampling on a toy-model, roughly displaying the features of the IILM. However, we have decided to leave out any orientation dependence because it is not necessarily a generic feature for the non-trivial backgrounds and would only introduce further complications in the Monte Carlo steps.

The simulations show that random sampling becomes very inefficient if the `temperature' is raised or the `mass' is lowered. We could run the toy-model for very long times and check that both the biased and unbiased simulations give equivalent results. For one particular set of parameters we actually found that ordinary Monte Carlo was not able to reach the correct equilibrium state; in this sense we lost ergodicity because the sample was not representative, having missed that region of phase space that corresponds to pair formation. We expect this to become a much more severe problem for the IILM because the higher dimensionality of the latter  will make it much harder to sample correctly the phase space volume leading to pair formation. Also, the simulations will be much more expensive, and long runs will be prohibitive.

We found that the parameters setting the bonding box have virtually no impact on the results, and that bulk properties agree on the $2\sigma$ level. This is as it should be because a strong dependence on these free parameters would introduce some new systematics and render the method impractical. We can therefore choose the parameters that give the smallest autocorrelation times.

Finally, we found that as the degree of dilution is increased, with fixed interaction strength, entropy-dominated configurations will eventually give the bulk contribution to the partition function, and the system will settle in a random state. The opposite equilibrium state, namely a highly correlated ensemble of pairs, is reached by increasing the interaction strength while holding the dilution parameter fixed; in that case the small entropy of these configurations is largely compensated by the gain in total energy, and the partition function is saturated by these energy-dominated states. These are generic features. In the IILM, however, the interactions become stronger and the system more dilute simultaneously as the temperature is increased. Whether the system ends up in a random ensemble or stays in the molecular phase after the chiral symmetry is restored depends on which quantity grows stronger and cannot be decided by the results presented in this study. It is, however, interesting to point to the possibility of a smaller topological susceptibility in the molecular phase as compared to the random phase. In light of the ultimate goal of this series of papers, this could lead to an axion mass that is rather different from the standard one computed in the dilute gas approximation.

\section*{Acknowledgements}

We are very grateful for many discussions with E.P.S. Shellard. Also, elucidating remarks by D. Kofke and B. Smit are gratefully acknowledged. Simulations were performed on the COSMOS supercomputer (an Altix 4700) which is funded by STFC, HEFCE and SGI. This work was supported by STFC grant PPA/S/S2004/03793 and an Isaac Newton Trust European Research Studentship.


\end{document}